\documentclass[12pt,a4paper]{article}

\usepackage{amsmath}
\usepackage{amssymb}
\usepackage{amsfonts}
\usepackage{graphicx}
\usepackage{subfig}
\usepackage{caption}
\usepackage{dcolumn}
\usepackage{bm}
\usepackage[
colorlinks,linkcolor=blue,citecolor=magenta]{hyperref}
\usepackage[top=3cm,right=2cm,bottom=2.5cm,left=2cm]{geometry}
\begin{document}
\title{A Generalized Model for the Classical Relativistic Spinning Particle}

\author{Mahdi Hajihashemi\thanks{mehdi.hajihashemi@ph.iut.ac.ir}, Ahmad Shirzad\thanks{shirzad@ipm.ir}\\
{\it Department of Physics, Isfahan University of Technology,}\\
{\it Isfahan 84156-83111, Iran }}
\maketitle
\begin{abstract}
Following the Poincare algebra, in the Hamiltonian approach, for a free spinning particle and
using the Casimirs of the algebra, we
construct systematically a set of Lagrangians for the relativistic
spinning particle which includes the Lagrangian given in the
literature. We analyze the dynamics of this generalized system in the Lagrangian formulation
and show that the equations of motion support an oscillatory solution corresponding to
the spinning nature of the  system. Then we analyze the
canonical structure of the system and present the correct gauge
suitable for the spinning motion of the system.
\end{abstract}
PASC number: 11.30.Cp
\section{Introduction}
There exist widespread efforts to construct a classical model for free
relativistic spinning particle (see for instance
references \cite{cor} and \cite{cor2} ). One of the earliest  examples belongs to Bargman,
Michel and Telegdi \cite{sl}. The next one was obtained by Hanson
and Regge \cite{hanson}. Subsequent models in this regard can be
seen in references \cite{ster}, \cite{ster2} and \cite{ster3}. Recently there are some works that interpret the so called  "relativistic Zitterbewegung effect" by a classical relativistic model for spinning
  particle \cite{Deri},\cite{deri1}. The last and more
sophisticated model is  proposed by Segal, Lyakhovich and Kuzenko
\cite{kuz} and is suggested independently by Staruszkiewicz \cite{7}.
The pioneer idea of Wigner \cite{qm}  that quantum mechanical
systems should be classified  according to irreducible
representations  of the Poincare group has a basic role in
construction of these models. As is well-known, representations of the
Poincare group are labeled by numerical values of two Casimirs of
the group, namely,
\begin{equation}
p^{2}=p{_\mu}p^{\mu},     \hspace{15mm}   W^{2}=W{_\mu}W^{\mu };~~
\label{a-1}
\end{equation}
where $W^\mu$ is the Pauli-Lubanski pseudovector defined as
\begin{equation}
W{_\mu}=-\frac{1}{2}\varepsilon_{\mu\nu\rho\sigma}M^{\nu\rho}p^{\sigma},
\label{a-2}
\end{equation}
in which $p^{\mu}$  are canonical momenta and $M^{\nu\rho}$ are the
angular momenta of the system. The Poisson brackets of $p^{\mu}$ and $M^{\nu\rho}$  satisfy the Poincare algebra. The
numerical values of $p^{2}$ and $W^{2}$, signify  the  mass and
spin of the system. For a spinless relativistic particle with
coordinates $x^\mu$ the Lagrangian (in natural units) is
\begin{equation}
L=-m \sqrt{\dot{x}^\mu \dot{x}_\mu}, \label{a-3}
\end{equation}
where $\dot{x}^\mu \equiv d x^\mu /d s$ and $s$ is a monotonic
function of the proper time. For this system we have
\begin{equation}
M_{\mu\nu }=x_\mu p_\nu -x_\nu p_\mu. \label{a-4}
 \end{equation}
One can easily verify that the pseudovector $W^{\mu }$ in Eq.
(\ref{a-2}) vanishes identically; hence the model exhibits no spin.

In order to describe a spinning particle in non-relativistic quantum
mechanics, one conventionally  enlarges the Hilbert space by adding
a two dimensional spinor space to the space of states. However, for describing a "classical relativistic spinning
particle"  the only way is enlarging the Minkowski space. For this reason in
Ref. \cite{7}  a light-like cut of a new Minkowski
space, coordinated by $ k_{\mu}$, is added to the ordinary
Minkowski space;  so that $W^{\mu }$ no longer vanishes. As
we will see, the model needs a new parameter $l$ with dimension of
length to describe the magnitude of spin. In Ref. \cite{la}, based
on an analysis in Lagrangian approach, it is shown that
when the parameter $l$ is equal to the Compton wavelength, the
magnitude of spin is $\frac{\hbar}{2}$.

One important problem in the framework of the Lagrangian
formulation is finding systematically a Lagrangian for the
spinning particle on the basis of the algebraic properties of the
Poincare group. This has not been done yet. Instead, a
complicated Lagrangian  is proposed which leads
to a non-vanishing $W^\mu$ vector. It is not, however, clear in
what sense it has been appeared and is it the unique one for the
corresponding space of dynamical variables or not.

Some interesting features also show up in the Hamiltonian formulation.
The constraint structure of the model is remarkable, and one needs
to fix the gauges due to the first class constraints of the
system. The problem is which gauge gives a suitable model
describing the relativistic spinning particle. In Ref.
\cite{ghosh}, a special gauge is proposed, and it is claimed that
it gives the well-established results of Ref. \cite{la} in the
Lagrangian approach.

 Our aim in this paper is to find solutions for both above
problems. First, by demanding that the Casimirs of the symmetry
group (here, the Poincare group) should appear in the Hamiltonian
formalism as first class constraints, we use the inverse of
Legendre transformation to build up the Lagrangian of the model.
We will do this for a spinless as well as a spinning particle in
section 2. In this way we "find" systematically  a large class of
the Lagrangians describing the spinning particle which include the
Lagrangian "proposed" in the literature as a special case.
In section 3 we analyze the Lagrangian dynamics of the system and suggest a
nontrivial solution for the equations of motion. Then we will evaluate the constraint structure of the model in
section 4 and find first class constraints as generators
of the gauge transformations. For fixing the gauge freedom, we propose a set of gauge fixing
conditions which  lead to expected results
of the Lagrangian analysis. Section 5
denotes to a summary and conclusions of the results. Throughout this paper
we assume the metric of the flat space is $g_{\mu\nu}$=diag(1,-1,-1,-1).

\section{  Relativistic spinning particle model }
Using the symmetries of the system, we try to
find the most general form of the Lagrangian of a relativistic
spinning particle. We introduce our systematic method
first for a spinless particle in Minkowski space. We see that it
leads to the famous Lagrangian of Eq. (\ref{a-3}). Then we will use the
method for the case of a spinning particle and find a generalized
Lagrangian which  reduce to the Lagrangian introduced in Ref. \cite{7} by a special choice of parameters.

A point particle is described relativistically by the
space-time coordinates $x^\mu$ which are functions of a
parameter $\tau'$ along the world-line, where $\tau'$ can be any monotonic
function of the proper time $\tau$. Every covariant theory
of the point particle should be insensible to the choice of the
evolution parameter. In other words, the suggested Lagrangian should be invariant under the change
\begin{equation}
\tau \rightarrow \tau'(\tau), \label{a-5}
\end{equation}
provided that $d\tau'/d\tau >0$. For the special
choice in which the evolution parameter is the proper time, we have
the gauge fixing condition $U^2=1$, where $U^\mu \equiv
dx^\mu/d\tau$ is the four-velocity of the particle. For an
arbitrary evolution parameter $\tau'(\tau)$ we have $U'^\mu \equiv
dx^\mu/d\tau'$ and the gauge fixing condition is
$U^2=(d\tau/d\tau')^2$.
Suppose $p_{\mu}$  is the momentum conjugate to the coordinate
$x^\mu$ with the following Poisson bracket
\begin{equation}
\{x_{\mu},p_{\nu}\}=g_{\mu \nu}. \label{a-6}
\end{equation}
However, the physical sector of
the phase space is restricted by constraints due to the
gauge symmetry. The generators of Lorentz group are
\begin{equation}
M_{\mu\nu}=x_{\mu}p_{\nu}-x_{\nu}p_{\mu}.\label{a-7}
\end{equation}
Using the fundamental Poisson brackets
(\ref{a-6}), it is easy to see $M_{\mu\nu}$ and $p_{\mu}$ satisfy
the Poincare algebra, as follows
 \begin{eqnarray} & &
 \{p_{\mu},p_{\nu}\}=0, \nonumber \\ &&
\{M_{\mu\nu},p_{\alpha}\}=g_{\mu\alpha}p_{\nu}-g_{\nu\alpha}p_{\mu},
\label{a-8} \\ && \{M_{\alpha\beta},M_{\mu\nu}\}= g_{\mu\beta}
M_{\nu\alpha} +g_{\nu\alpha}M_{\mu\beta}+ g_{\mu\alpha}
M_{\beta\nu}+ g_{\nu\beta}M_{\alpha\mu}. \nonumber
 \end{eqnarray}
Clearly the first Casimir $p^\mu p_\mu$ is nontrivial and can be
identified, in natural units ($\hbar=c=1)$, as the mass squared of
the point particle. Hence, the dynamics of the system in
Hamiltonian framework should be constructed such that the
constraint
\begin{equation}
p^{2}-m^{2}=0 \label{a-9}
\end{equation}
holds identically on the physical subspace of the phase space.
For a spinless free particle the constraint (\ref{a-9}) is the
only requirement to be considered and there is no other parameter
or quantum number associated to the particle. On the other hand,
for a system restricted to the eight phase space coordinates
$x^\mu$ and $p_\mu$ the second Casimir defined by Eq. (\ref{a-2})
vanishes identically. So there is no spin for a particle described
by the phase space coordinates $(x^\mu,p_\mu)$.

Now we want to find (and not suggest) the Lagrangian of the free
spinless particle assuming that the system  has reparametrization
symmetry under the transformation (\ref{a-5}), and obeys
the constraint (\ref{a-9}) in the Hamiltonian formulation.

It is well-known that for a point particle
where space-time coordinates are considered as functions of an
arbitrary evolution parameter $\tau$, the canonical Hamiltonian
vanishes (see section 8 of Ref. \cite{gold}). Hence, the total
Hamiltonian, which is responsible for the dynamics of the system
in the framework of the Dirac constraint theory \cite{dir}, is
constructed with the only constraint of the system as
\begin{equation}
H=\frac{e}{2}(p^{2}-m^{2})\label{a-10}
\end{equation}
where $e$ is the Lagrange multiplier recognized as the ein-bein
parameter in the literature. Using $\dot{x^{\mu}} =\{x^{\mu},
H\}=ep^{\mu}$, and Legendre transformation from the Hamiltonian to the
Lagrangian formulation we have
\begin{equation}
L=\dot{x^{\mu}}p_{\mu}-H =\frac{\dot{x}^{2}}{2e}+\frac{e}{2}m^{2}.
\label{a-12}
\end{equation}
The last statement is the  Polyakov form of the Lagrangian of a  spinless particle. The ein-bein parameter appears as
an auxiliary variable in the Lagrangian (\ref{a-12}).
Equation of motion of $e$ gives
$ e=-\sqrt{\dot{x}^{2}}/m$. By substituting this result in the
Lagrangian (\ref{a-12}) we obtain the Nambu-Goto form of the
Lagrangian as $L= -m \sqrt{\dot{x}^2}$.

Let us generalize the above method to a spinning free particle.
Clearly we should extend the coordinate space such that the second
Casimir of Eqs. (\ref{a-1}) does not vanish any more. So, we
should have a second constraint in the framework of the Dirac theory. In this way
we can construct the total Hamiltonian as a linear combination of
the constraints by using two different ein-bein variables.
Performing a Legendre transformation from the Hamiltonian to
the Lagrangian formulation finally brings us to the Polyakov form of
the Lagrangian of a relativistic spinning particle. This leads to
the Nambo-Goto form of the Lagrangian upon eliminating the
auxiliary ein-bein variables. Let us show the details in the following.

In order to  extend the ordinary Minkowski space we ascribe an additional four vector $k^{\mu}$ to the
particle. The enlarged phase space contains ($x^{\mu}$,$p_{\mu}$)
as well as ($k^{\mu}$,$q_{\mu}$) as canonical variables,
where $q_{\mu}$ is the conjugate momentum to $k^{\mu}$.
 The fundamental non vanishing Poison brackets in this phase space are
\begin{equation}
\{k_{\mu},q_{\nu}\}=g_{\mu\nu},~~~~~~~~~~~~~~
\{x_{\mu},p_{\nu}\}=g_{\mu\nu}.
\end{equation}
In the enlarged phase space, translation generators are still
$p_{\mu}$ while Lorentz  generators are
\begin{equation}
M_{\mu\nu}=x_{\mu}p_{\nu}-x_{\nu}p_{\mu}+k_{\mu}q_{\nu}-k_{\nu}q_{\mu}.
\label{a-13}
\end{equation}
One can easily check that $p_{\mu}$ and $M_{\mu\nu}$ of Eq.(\ref{a-13})
satisfy the Poincare algebra (\ref{a-8}). In the enlarged  phase space $W^{2}$ is no
more vanishing, instead we have
 \begin{equation}
W^{2}=-k^{2}q^{2}p^{2}+k^{2}(p.q)^{2}-2(p.k)(p.q)(k.q)+(k.q)^{2}p^{2}+(p.k)^{2}q^{2}.
\end{equation}

Suppose $k^{\mu}$ has arbitrary physical dimension.
In all physical quantities $k^{\mu}$ appears  multiplied by  its conjugate momentum $q_{\mu}$.
Hence, the physical dimension of  $k^{\mu}$ has  no influence on the dimension of the corresponding quantity.
The quantity $W^{2}$ has the dimension of mass squared. The simplest choice is to
assume that $W^2$ is proportional to $m^{4}l^{2}$, where the new
parameter $l$, identifying the magnitude of spin, has dimension of
length. If we put $W^{2}+ C m^{4}l^{2}\approx 0$  for a
dimensionless parameter $C$, we should have the following
constraint on the phase space variables,
 \begin{equation}
-k^{2}q^{2}p^{2}+k^{2}(p.q)^{2}-2(p.k)(p.q)(k.q)+(k.q)^{2}p^{2}+(p.k)^{2}q^{2}+\ C m^{4}l^{2}
\approx 0,
 \end{equation}
 where the symbol $\approx $ means weak equality.
 Remembering the constraint $p^2-m^2 \approx 0$, and introducing
 two ein-bein variables $e_1$ and $e_2$, the total Hamiltonian of the system is
 \begin{equation}
H=\frac{e_{1}}{2}(p^{2}-m^{2})+\frac{e_{2}}{2}\big(-k^{2}q^{2}p^{2}+k^{2}(p.q)^{2}-2(p.k)(p.q)(k.q)+(k.q)^{2}p^{2}+(p.k)^{2}q^{2}+C m^{4}l^{2}\big).
\label{H}
 \end{equation}
  Using  this suggested Hamiltonian, the
time derivative of $k.q$ and $k^{2}$ read
  \begin{equation}
\frac{d }{d\tau}(k.q)=\{k.q ,H\}=0, ~~~~~~~~~~~~~~~~~~~\frac{d }{d\tau}(k^{2})=\{k^{2} ,H\}=0,
\end{equation}
 which means that $k.q$ and $k^{2}$  are constants of motion; so we can replace them in Eq. (\ref{H}) by  constant values, i.e.  $k.q=\alpha$ and $k^{2}=\beta$.
Notice that $k.q$ is dimensionless while $k^{2}$  has dimension of k squared. Since $m$ and $l$ are assumed as the main physical
parameters of the  spinning point  particle, there is no reason to consider additional physical parameters as the value of $k^{2}$. Hence, the
most natural choice for $\beta$ is zero. In fact in the literature  \cite{7,la,ghosh}, it is conventional to assume  $k^{\mu}$  as a light-like vector.
However, the above analysis shows that otherwise, one needs to describe
the spinning point particle with three physical parameters. However, in the real
world every spinning particle is described by just two parameters $m$  and $\hbar$ (or alternatively by $m$ and $l$).

 In this way, taking $k^{2}=0$ and $k.q=\alpha$, the Hamiltonian (\ref{H}) becomes
  \begin{equation}
H=\frac{e_{1}}{2}(p^{2}-m^{2})+\frac{e_{2}}{2}\big(-2\alpha(p.k)(p.q)+\alpha^{2}p^{2}+(p.k)^{2}q^{2}+C m^{4}l^{2}\big).
\label{H1}
 \end{equation}
We are allowed to work with the Hamiltonian (\ref{H1}) provided that the constraints $k^2=0$
and $k.q=\alpha$ are somehow satisfied in the dynamics of the theory. We will come back to this point once again.
From the Hamiltonian (\ref{H1})  the time derivatives  of $k_{\mu}$ and $x_{\mu}$  are as follows
 \begin{eqnarray} &&
\dot{k}_{\mu}=e_{2}(p.k)^{2}q_{\mu}
- e_{2}\alpha(p.k)p_{\mu},
\nonumber\label{dt1}
\\ &&\dot{x}_{\mu}=e_{1}p_{\mu}
-e_{2}\alpha(p.q)
k_{\mu}-e_{2}\alpha (p.k) q_{\mu}+ e_{2} \alpha^{2} p_{\mu}+ e_{2}(p.k)q^{2}k_{\mu}.\label{dt2}
 \end{eqnarray}
 By Legendre transformation (in the reverse direction) as
$L=\dot{x^{\mu}}p_{\mu}+ \dot{k^{\mu}}q_{\mu}-H$, we obtain
  \begin{equation} L= \frac{e_{1}} {2}p^{2}
-2e_{2}\alpha(q.p)(p.k)+
+\frac{3}{2}e_{2}(p.k)^{2}q^{2}
 + \frac{1}{2}e_{2}
\alpha^{2}p^{2}+\frac{e_{1}}{2}m^{2}-\frac{1}{2}e_{2}C
m^{4}l^{2}.\label{pol}
 \end{equation}
 To this end we should transfer from the momentum variables $p_\mu$
and $q_\mu$ to velocities $\dot{x}_{\mu}$ and $\dot{k}_{\mu}$ in
the expression of the Lagrangian (\ref{pol}).
Using Eqs. (\ref{dt1}), we have
   \begin{eqnarray} &&  k.\dot{x}=e_{1}(p.k),\nonumber \\&& \dot{k}^{2}
=e_{2}^{2}(p.k)^{4}q^{2}+e_{2}^{2}\alpha^{2}(p.k)^{2}p^{2}
-2\alpha e_{2}^{2}(p.k)^{3}(p.q),\nonumber \\ &&
\dot{k}.\dot{x}=2\alpha^{2}e_{2}^{2}(p.k)^{2}(p.q)
+e_{1}e_{2}(p.k)^{2}(p.q) -\alpha
e_{1}e_{2}p^{2}(p.k)-\alpha^{3}e_{2}^{2}p^{2}(p.k)-\alpha
e_{2}^{2}q^{2}(p.k)^{3},  \nonumber \\ &&
\dot{x}^{2}=-4e_{1}e_{2}\alpha(p.k)(p.q)+e_{1}^{2}p^{2}+
2e_{1}e_{2}(p.k)^{2}q^{2} \label{s-100} \\ && \hspace{5cm}
+2e_{1}e_{2}\alpha^{2}p^{2}+
\alpha^{2}e_{2}^{2}(p.k)^{2}q^{2}+\alpha^{4}e_{2}^{2}p^{2}
-2\alpha^{3}e_{2}^{2}(p.k)(p.q).\nonumber
 \end{eqnarray}
 One can  invert the Eqs. (\ref{s-100}) to obtain $p^{2}$ , $q^{2}$
, $p.q$ and $ p.k$ in terms of $ \dot{x}^{2}$ ,  $ \dot{k}^{2}$,
$k.\dot{k}$ and $\dot{k}.\dot{x}$ as follows

 \begin{eqnarray} &&
p.k= \frac{k.\dot{x}}{e_{1}}, \nonumber \\
&& p^{2}=\frac{(k.\dot{x})^{2}\dot{x}^{2}e_{2}-2\dot{k}^{2}
e_{1}^{3}- \dot{k}^{2} \alpha^{2}e_{1}^{2}e_{2}}
{(k.\dot{x})^{2}e_{1}^{2}e_{2}}, \nonumber \\ && p.q=\frac{k.\dot{x}
(\dot{k}.\dot{x}) e_{1}^{2}+(k.\dot{x})^{2}\dot{x}^{2}\alpha
e_{2}-\dot{k}^{2} \alpha e_{1}^{3}
-\dot{k}^{2}\alpha^{3}e^{2}e_{1}} {(k.\dot{x})^{3} e_{1}e_{2}},
\label{s-101} \\ && q^{2}=\frac{\dot{k}^{2}e_{1}^{4}
+2k.\dot{x}(\dot{k}.\dot{x})\alpha
e_{1}^{2}e_{2}+(k.\dot{x})^{2}\dot{x}^{2}\alpha^{2}
e_{2}^{2}-\dot{k}^{2}\alpha^{4}e_{1}^{2}e_{2}^{2}}
{(k.\dot{x})^{4}e_{2}^{2}}. \nonumber
 \end{eqnarray}
Substituting these expression in Eq.(\ref{pol}), the Polyakov form of the
Lagrangian appears as
  \begin{equation}
L=e_{1}\frac{m^{2}}{2}-\frac{1}{2}C e_{2}m^{4}l^{2}+ \frac{\dot{x}^{2}}
{2e_{1}}+\frac{e_{1}^{2}\dot{k}^{2}} {2e_{2}(k.\dot{x})^{2}}+\frac{\dot{k}^{2}
\alpha^{2}e_{1}}{2(k.\dot{x})^{2}}+\frac{(\dot{k}.\dot{x})
\alpha}{(k.\dot{x})}.\label{poll}
  \end{equation}
The auxiliary variables $e_{1}$ and $e_{2}$ can be obtained in
terms of the dynamical variables of the spinning particle by using
the corresponding equations of motion as follows
 \begin{eqnarray} &&
e_{1}=\frac{-l\sqrt{\dot{x}^2}}{\sqrt{\frac{2l^{3}m^{2}\sqrt{-\dot{k}^{2}}}
{k.\dot{x}}  \sqrt{C}+
\frac{l^{2}\alpha^{2}\dot{k}^{2}}{(k.\dot{x})^{2}}+l^{2}m^{2}}} ,\label{e1}\\
&& e_{2}=\frac{\frac{l\sqrt{-\dot{k}^{2}}}
{k.\dot{x}}\sqrt{\dot{x}^2}} {m^{2}l\sqrt{C}\sqrt{\frac{2l^{3}m^{2}
\sqrt{-\dot{k}^{2}}}{k.\dot{x}} \sqrt{C}+
\frac{l^{2}\alpha^{2}\dot{k}^{2}}{(k.\dot{x})^{2}}+l^{2}m^{2}}} .\label{e2}
 \end{eqnarray}
Substituting $e_{1}$ and $e_{2}$  from Eqs. (\ref{e1})  and (\ref{e2}) in the Polyakov form
of the Lagrangian (\ref{poll}), we obtain the Nambu-Goto form of Lagrangian as
\begin{equation}
L=-\frac{\sqrt{\dot{x}^{2}}}{l}\sqrt{\frac{2l^{3}m^{2}
		\sqrt{-\dot{k}^{2}}}{k.\dot{x}} \sqrt{C}+
	\frac{l^{2}\alpha^{2}\dot{k}^{2}}{(k.\dot{x})^{2}}+l^{2}m^{2}}+ \frac{\alpha (\dot{k}.\dot{x})}{k.\dot{x}}.
\end{equation}
As we mentioned after Eq. (\ref{H1}), we need to guarantee the constraints $k.q=\alpha$ and $k^2=0$ in order to get the correct form of the Casimir constraints. As we will see in the following sections, the constraint $k.q=\alpha$ emerges naturally during the dynamical procedure. However, we need to impose the constraint $k^2=0$ by using a Lagrange multiplier in the Lagrangian. The final result is
\begin{equation}
L=-\frac{\sqrt{\dot{x}^{2}}}{l}\sqrt{\frac{2l^{3}m^{2}
\sqrt{-\dot{k}^{2}}}{k.\dot{x}} \sqrt{C}+
\frac{l^{2}\alpha^{2}\dot{k}^{2}}{(k.\dot{x})^{2}}+l^{2}m^{2}}+ \frac{\alpha (\dot{k}.\dot{x})}{k.\dot{x}}-\lambda
k^2.\label{Lag}
 \end{equation}

This is  the most general form of the Lagrangian of a classically
spinning particle. By choosing $C=1/4$ and $\alpha=0$ Eq.
(\ref{Lag}) reduce to the action proposed in Ref. \cite{7}. Hence
the proposed Lagrangian is not unique and should be considered as
a special case of a large class of Lagrangians given in Eq.
(\ref{Lag}).
\section{Lagrangian dynamic}
In this section we try to find a nontrivial solution for the Euler-Lagrange equations of motion of the model
obtained in the previous section.  Suppose we are given from
the very beginning the action
\begin{equation}
S=\int d\tau  \left( -\frac{\sqrt{\dot{x}^{2}}}{l}\sqrt{\frac{2l^{3}m^{2}
\sqrt{-\dot{k}^{2}}}{k.\dot{x}} \sqrt{C}+
\frac{l^{2}\alpha^{2}\dot{k}^{2}}{(k.\dot{x})^{2}}+l^{2}m^{2}} +  \frac{\alpha (\dot{k}.\dot{x})}{k.\dot{x}} -
 \lambda k^2 \right) ,
 \end{equation}
 where $\tau$ is the proper time, $m$  is  the mass and $l$ is a
parameter with the dimension of length that identifies  the magnitude
of spin. The particle
is described by the dynamical variables $(x_\mu,k_\mu)$  while $\lambda$ is the Lagrange multiplier
corresponding to the assumption that $k$ is light-like.  The equations of motion
for the variables $x^{\mu}$ and $k^{\mu}$ are as follows
\begin{eqnarray}
\frac{\partial p_{\mu}}{\partial \tau}=0,\label{eqx}~~~~~~~~~~~~~~~~~~~~~~~~~~~~~~~~~~~~~~~~~~~~~~~~~~~~~~~~~~~~~~~~~~~~
 \end{eqnarray}
\begin{eqnarray}
\frac{\partial q_{\mu}}{\partial \tau}
=\frac{\sqrt{\dot{x}^{2}}(\frac{2 l^{3}\sqrt{C}m^{2}\sqrt{-\dot{k}^{2}}\dot{x}_{\mu}}{(k.\dot{x})^{2}}
+\frac{2 l^{2}\alpha^{2}\dot{k}^{2}\dot{x}_{\mu} }{(k.\dot{x})^{3}})}{2l\sqrt{\frac{2l^{3}m^{2}
\sqrt{-\dot{k}^{2}}}{k.\dot{x}} \sqrt{C}+
\frac{l^{2}\alpha^{2}\dot{k}^{2}}{(k.\dot{x})^{2}}+l^{2}m^{2}}} -\frac{\alpha (\dot{k}.\dot{x})\dot{x}_{\mu}}{(k.\dot{x})^{2}}-2 \lambda k_{\mu},\label{eqk}
 \end{eqnarray}
where
\begin{eqnarray}
p_{\mu}=-\frac{\sqrt{\frac{2l^{3}m^{2}
\sqrt{-\dot{k}^{2}}}{k.\dot{x}} \sqrt{C}+
\frac{l^{2}\alpha^{2}\dot{k}^{2}}{(k.\dot{x})^{2}}+l^{2}m^{2}} \dot{x_{\mu}}}{l \sqrt{\dot{x}^{2}}}
+\frac{\sqrt{\dot{x}^{2}}(\frac{2 l^{3}\sqrt{C}m^{2}\sqrt{-\dot{k}^{2}}k_{\mu}}{(k.\dot{x})^{2}}
+\frac{2 l^{2}\alpha^{2}\dot{k}^{2}k_{\mu} }{(k.\dot{x})^{3}})}{2l\sqrt{\frac{2l^{3}m^{2}
\sqrt{-\dot{k}^{2}}}{k.\dot{x}} \sqrt{C}+
\frac{l^{2}\alpha^{2}\dot{k}^{2}}{(k.\dot{x})^{2}}+l^{2}m^{2}}}
+\frac{\alpha \dot{k}_{\mu}}{k.\dot{x}}-\frac{\alpha(\dot{k}.\dot{x})k_{\mu}}{(k.\dot{x})^{2}},\label{mo1}
\end{eqnarray}
and
\begin{equation}
q_{\mu}=\frac{\sqrt{\dot{x}^{2}}(\frac{2 l^{3}\sqrt{C}m^{2}k_{\mu}}{(k.\dot{x}) \sqrt{-\dot{k}^{2}}}
-\frac{2 l^{2}\alpha^{2}\dot{k}_{\mu} }{(k.\dot{x})^{2}})}{2l\sqrt{\frac{2l^{3}m^{2}
\sqrt{-\dot{k}^{2}}}{k.\dot{x}} \sqrt{C}+
\frac{l^{2}\alpha^{2}\dot{k}^{2}}{(k.\dot{x})^{2}}+l^{2}m^{2}}} +\frac{\alpha \dot{x}_{\mu}}{k.\dot{x}}. \label{mo2}
\end{equation}

As we said before, the  Lagrangian (\ref{Lag}) reduce to the Lagrangian suggested in Ref. \cite{7} by  choosing $C=1/4$ and $\alpha=0$.
In Ref. \cite{la} the dynamics of this special case is analyzed and the authors show  that a nontrivial oscillatory solution
 can be proposed for the system. Hence, we expect to find a more general oscillatory solution for the equations of motion (\ref{eqx}) and (\ref{eqk}).

 By direct calculation one can show that the following oscillatory solution satisfies  the equations of motion (\ref{eqx}) and (\ref{eqk}),
 \begin{eqnarray} && x_{\mu}=(\tau,l\sqrt{C}\cos(\omega\tau)-\frac{\alpha}{m}\sin(\omega\tau)
,-l\sqrt{C}\sin(\omega\tau)-\frac{\alpha}{m}\cos(\omega\tau),0)\nonumber\\
&&k_{\mu}=(-\gamma,-\gamma\sin(\omega\tau),- \gamma \cos(\omega
\tau),0), \label{kk}
 \end{eqnarray}
where $\gamma$ and $\omega$ are   numerical constant. For the solution (\ref{kk})  to be valid the Lagrange multiplier  $\lambda$ should be taken as
\begin{equation}
\lambda=-\frac{ml\omega\sqrt{C}}{2\gamma^{2}(1+\sqrt{C}l\omega)}.
 \end{equation}
To verify that the  solution (\ref{kk}) satisfies  the equations of motion (\ref{eqx}) and (\ref{eqk}) one needs to notice that for this solution the quantities    $k.\dot{x}$, $\dot{x}^{2}$, $\dot{k}^{2}$ and $\dot{k}.\dot{x}$  are constants as follows
\begin{eqnarray}&&\dot{k}^{2}=-\gamma^{2}\omega^{2},\\&&
\dot{k}.\dot{x}=-\frac{\alpha\gamma\omega^{2}}{m}\nonumber,\\&&
k.\dot{x}=-\gamma(1+l\omega\sqrt{C})\nonumber,\\&&
\dot{x}^{2}=1-Cl^{2}\omega^{2}-\frac{\alpha^{2}\omega^{2}}{m^{2}}.\nonumber
\end{eqnarray}

\section{Hamiltonian dynamic}

\subsection{constraint structure}

In this section we investigate the constraint structure of our
spinning point particle. It is expected that
constraints that we used to guess the Hamiltonian in
section 2 appear again in what follows. Beginning from  the Lagrangian (\ref{Lag})
one can directly define the conjugate momenta as given in Eqs (\ref{mo1}) and (\ref{mo2}) in addition to
\begin{equation}
P_{\lambda}=0 \label{mo3}
\end{equation}

(where $P_{\lambda}$ is conjugate momenta of parameter $\lambda$). From Eq. (\ref{mo3}) one primary constraint of the system is
$\varphi_{1}\equiv P_{\lambda}\approx 0$. Since the Lagrangian
(\ref{Lag}) contains only homogeneous function of first degree
with respect to the velocities, (i.e. $L \rightarrow rL$ under the
scaling $(\dot x,\dot q)\rightarrow (r\dot x,r\dot q)$), the
canonical Hamiltonian includes just the last term of the
Lagrangian as
\begin{equation}
H_{c}=-\lambda k^{2}.
 \end{equation}
Consistency of the primary constraint
$\varphi_{1}$ under time translation yields $\varphi_{2}\equiv
k^{2}\approx 0$ which recalls us that $k_{\mu}$ is light-like.
Consistency of $\varphi_{2}$  under time translation do not yield
a new constraint. However,  we have two more constraints by
manipulating  the $q_{\mu}$ and $p_{\mu}$ in (\ref{mo1}) and
(\ref{mo2}) as follows
 \begin{equation}
\varphi_{3}\equiv  p^{2}-m^{2}\approx 0,~~~~~~~ \varphi_{4} \equiv
-2\alpha(p.k)(p.q)+\alpha^{2}m^{2}+(p.k)^{2}q^{2}+\sqrt{C}m^{4}l^{2}\approx 0.\label{gh}
\end{equation}

Consistency of $\varphi_{4}$ under time translation yields $
\lambda (p.k)^{2}(k.q)-\lambda(p.k)^{2} \approx 0.$ One can deduce from the second
equation of (\ref{gh}) that $ p.k\neq0$. Therefore $\varphi_{5}\equiv
k.q-\alpha \approx0$ is another constraint of the system. This constraint is exactly what we expected to find from the dynamics of the theory when we reduced the Hamiltonian (\ref{H}) to the Hamiltonian (\ref{H1}).
Consistency of
$\varphi_{3}$ and  $\varphi_{5}$ under time translation do not
yield  any new constraint. Since non of the Lagrange multipliers
is determined in the process of consistency, all of the five
constraints obtained are first class. This can be directly
verified by calculating the Poisson brackets of the constraints.

One can directly verify that on the constraint surface we have
\begin{equation}
p^\mu p_\mu =m^2~;~~W^\mu W_\mu =-\sqrt{C}m^{4}l^{2}
\end{equation}
where $W_\mu$ defined in Eq. (\ref{a-2}) is constructed on the
basis of momenta derived in Eqs. (\ref{mo1}) and (\ref{mo2}). As
mentioned earlier, the magnitude of the obtained Casimirs are
related to the mass and spin of the particle.

As expected, the constraints we used for obtaining the action
(\ref{Lag}), appear  again in the constraint structure of the
system. The constraints $\varphi_{3}$ and $\varphi_{4}$ are
Casimirs of the spinning point particle. The constraint
$\varphi_{2}$ is the same condition we assumed in
the previous section that the vector $k^\mu$ is light-like. The
constraint $\varphi_{5}$  corresponds to our assumption
about the value of  $k.q$  in section 2. Finally the primary constraint
$\varphi_{1}$ is due to considering the
dynamic-less Lagrange multiplier $\lambda$.

\subsection{Gauge fixing process}
In the previous subsection we found that the  model has five first class constraints and the
canonical Hamiltonian $H_{c}=-\lambda k^{2}$ vanish on the constraint
surface. As explained in Ref. \cite{dir} in such systems, the dynamics
is governed by the  extended Hamiltonian which is a linear
combination of  the first class constraints. In our case we have
 \begin{equation}
H_{E}= C_{1} (P_{\lambda})+ C_{2}(k^{2}) + C_{3} (p^{2}-m^{2})
+C_{4} \bigg(-2\alpha(p.k)(p.q)+\alpha^{2}m^{2}+(p.k)^{2}q^{2}+\sqrt{C}m^{4}l^{2}\bigg)+C_{5}(k.q).
\end{equation}
The multipliers $C_{i}$ should  be fixed by gauge fixing process.
In this process one imposes the gauge fixing conditions (GFC)
\cite{dr shirzad} $\psi_{g}^{i}$ corresponding to the first class
constraints $\varphi_{i}$ such that
 \begin{equation}
det\{\psi_{g}^{i},\varphi_{j}\} \neq 0.~~~~~~~~i,j=1,...5 \label{x1}
 \end{equation}
It is clear that this process is not unique. In the current
problem we wish to find an appropriate set of GFC's in the center
of mass frame (where $p_{\mu}=(m,0,0,0)$) that lead to the oscillatory
results obtained in Lagrangian analysis of section 3, i.e.
 \begin{eqnarray} && x_{\mu}=(\tau,l\sqrt{C}\cos(\omega\tau)-\frac{\alpha}{m}\sin(\omega\tau)
,-l\sqrt{C}\sin(\omega\tau)-\frac{\alpha}{m}\cos(\omega\tau),0),\nonumber\\
&&k_{\mu}=(-\gamma,-\gamma\sin(\omega\tau),- \gamma \cos(\omega
\tau),0). \label{33} .
 \end{eqnarray}
 We suggest five GFC's as follows
  \begin{eqnarray} && \psi_{g}^{1}\sim  \lambda-1,\nonumber \\ &&
\psi_{g}^{2} \sim q_{2}  -\frac{ml\sqrt{C}}{\gamma}\sin(\omega\tau),\label{gfc}
\nonumber \\ && \psi_{g}^{3}\sim   x_{0}-\tau, \label{gfc2}\\
&& \psi_{g}^{4}\sim q_{1}  +\frac{ml\sqrt{C}}{\gamma}\cos(\omega\tau),
\nonumber  \\ && \psi_{g}^{5}\sim  q_{3} \nonumber
 \end{eqnarray}
where $\omega$ and $\gamma$ are nonzero numerical constants. Explicit calculation shows that the suggested GFC's satisfy the condition (\ref{x1}), i.e. the whole system of GFC's and the constraints constitute a system of second class constraints. Hence, we can determine
the multipliers $C_{i}$  in the extended Hamiltonian by
consistency of GFC's  via the relations
  \begin{equation}
\{\psi_{g}^{i},H_{E}\}+\frac{\partial \psi_{g}^{i}
}{\partial\tau}=0 \label{fs}.
 \end{equation}
In this way for consistency of $\psi_{g}^{1}$ to $\psi_{g}^{5}$ we
find respectively
 \begin{eqnarray} && C_{1}\{P_{\lambda},\lambda\} =0, \label{so1} \\
&& -2C_{4}q^{2}(p.k)p_{2}-C_{5}q_{2}-2C_{2}k_{2}+2C_{4}\alpha(p.q)p_{2}- \frac{m l \sqrt{C} \omega}{\gamma}
 \cos(\omega\tau)=0,\label{so2} \\ && 2C_{3}p_{0}+2C_{4}
q^{2}(p.k)k_{0}-2C_{4}\alpha(p.k)q_{0}-2C_{4}\alpha(p.q)k_{0}-1= 0,\label{so3} \\ &&
-2C_{4}q^{2}(p.k)p_{1}-C_{5}q_{1}-2C_{2}k_{1}+2C_{4}\alpha(p.q)p_{1}- \frac{m l \sqrt{C} \omega}{\gamma}
 \sin(\omega\tau)=0,\label{so4}\\ &&
-2C_{4}q^{2}(p.k)p_{3}-C_{5}q_{3}-2C_{2}k_{3}+2C_{4}\alpha(p.q)p_{3}=0.
\label{so5}
 \end{eqnarray}
  Eqs. (\ref{so1}-\ref{so5}) uniquely determine the coefficient $C_{1}-C_{5}$. This shows that five GFC's
  given in Eq. (\ref{gfc}) in fact fix the gauge completely. Except $C_1$ which is zero from Eq. (\ref{so1}), we find some
complicated results for $C_2$ to $C_5$ from Eqs.
(\ref{so2}-\ref{so5}). Note that we are not interested in the most
general solution of the equations of motion. We are just looking
for a special spinning solution obtained in the  Lagrangian dynamics. So we will be
happy if the equations of motion support such a solution.
Let us first investigate how many dynamical variables remain to be
determined after imposing the whole set of constraints and gauge
fixing conditions.

Here, we began with 18 variables (i.e. $x^\mu$, $p_\mu$, $k^\mu$,
$q_\mu$, $\lambda$ and $p_\lambda$) and continued with 5 first
class constraints and 5 gauge fixing conditions. There remain 8
degrees of freedom which are dynamical, that is, they obey first
order differential equations of motion and their values at any
given time depend on their initial values (or constants of
motion). Since there is no $x^{\mu}$ dependence in the constraints
and the assumed gauge affects only on the values of $C_i$, we have
$\{ p_i, H_E\}= 0$. So three components of the momentum are
dynamical through dynamical equations $\dot p_i=0$, regardless of
the gauge considered. Hence, we can choose a solution in which
three constant components of the momentum are zero (i.e. we go to the
center of mass reference frame).

Among the other variables, $p_0$ is fixed as $m$ due to the
constraint $\phi_3$, $p_\lambda$ vanishes due to the constraint
$\phi_1$ and $\lambda$, $x^0$ and $q_i$ are fixed by the gauge
considered in Eqs. (\ref{gfc2}). Among the 8 remaining variables, say $x^i$,
$q_0$ and $k^\mu$, 5 ones are dynamical and 3 ones should be derived
from the unused constraints $\phi_2$, $\phi_4$ and $\phi_5$. We
assume the dynamical variables are $x^i$, $k^1$ and $k^2$. Let us see whether the
dynamical equations support a solution in which $k^1=-k_1=\gamma \cos
\omega \tau$ and $k^2=-k_2=\gamma \sin \omega \tau$. With this choice Eq. (\ref{so5}) gives $k_3=0$ and the
constraint $\phi_2$ gives $k^0=-\gamma$. Then using the  dynamical equation $\dot k_i= \{k_i,H_E\}$  gives $q^0=-\frac{\alpha}{\gamma}$
 . Direct calculation using
Eqs. (\ref{so2}-\ref{so5}) and consistency of the assumed solution with the
dynamical equations $\dot k_i= \{k_i,H_E\}$ and $\dot q_i= \{q_i,H_E\}$ then gives the
coefficients $C_2$ to $C_5$ as follows
 \begin{equation}
C_{2}=\frac{\sqrt{C}ml\omega}{2\gamma^{2}},~~~~  C_{3}
=\frac{1+l\sqrt{C}\omega+\frac{\alpha^{2}\omega}{\sqrt{C}l m^{2}}}{2m},~~~~C_{4}=\frac{\omega}{2\sqrt{C}m^{3}l},
~~~~~~C_{5}=0.\label{anzats}
 \end{equation}
The extended Hamiltonian  by  using the above ansatz read
 \begin{equation}
H_{E}=\frac{1+l\sqrt{C}\omega+\frac{\alpha^{2}\omega}{\sqrt{C}l m^{2}}}{2m}(p^{2}-m^{2})+\frac{\omega \big(-2\alpha(p.k)(p.q)+\alpha^{2}m^{2}+(p.k)^{2}q^{2}+\sqrt{C}m^{4}l^{2}\big)}
{2\sqrt{C}m^{3}l}+\frac{\sqrt{C}ml\omega}{2\gamma^{2}}k^{2}.
 \end{equation}
Hence, we find the following equations of motion for the canonical
variables
 \begin{eqnarray} && \dot{x}_{\mu}=\frac{p_{\mu}}{m}+\frac{l\sqrt{C}\omega}{m}p_{\mu}+\frac{\alpha^{2} \omega}{\sqrt{C}l m^{2}}p_{\mu}+\frac{\alpha\omega\gamma}{\sqrt{C}l m^{2}}q_{\mu}+\frac{l\omega\sqrt{C}}{\gamma}k_{\mu},\\
&&\dot{k}_{\mu}=\frac{\gamma^{2}\omega}{m l \sqrt{C}}q_{\mu}+\frac{\alpha \omega \gamma}{\sqrt{C}l m^{2}}p_{\mu}, \\&&
\dot{q}_{\mu}=-\frac{\sqrt{C}l m\omega}{\gamma^{2}}k_{\mu}-\frac{l \omega \sqrt{C}}{\gamma}p_{\mu},\\&&\dot{p}_{\mu}=0.
 \end{eqnarray}
Now it is a straightforward exercise to see that the following
solution satisfies the above equations of motion,
 \begin{eqnarray} && x_{\mu}=(\tau,l\sqrt{C}\cos(\omega\tau)-\frac{\alpha}{m}\sin(\omega\tau)
,-l\sqrt{C}\sin(\omega\tau)-\frac{\alpha}{m}\cos(\omega\tau),0)\label{x},\\ && p_{\mu}=(m,0,0,0),\\
&&k_{\mu}=(-\gamma,-\gamma\sin(\omega\tau),- \gamma \cos(\omega
\tau),0), \label{k} \\ && q_{\mu}=(-\frac{\alpha}{\gamma},-\frac{m l \sqrt{C}} {\gamma}\cos
(\omega\tau), \frac{m l\sqrt{C} }{\gamma}\sin(\omega\tau),0)\label{q}.
 \end{eqnarray}
This solution is consistent with the GFC's given in Eqs.
(\ref{gfc2}) as well as our assumed ansatz for $p_i$, $k^1$ and
$k^2$. To be honest, we have managed the GFC's to be consistent
with the oscillatory solution of Eqs. (\ref{x}-\ref{q}). However,
the point is that the model allows us to construct such a spinning
solution.

This analysis shows that if we choose appropriate gauge fixing conditions
as given in Eqs. (\ref{gfc2}), we can revive the oscillatory mode
that appeared in the Lagrangian analysis.

It may be wonderful in the first look why in the center of mass
frame the particle goes through a circular trajectory. However, as
mentioned earlier, the structure of the model is such that the space
part of the Pauli-Lubanski pseudo-vector identifies the spin vector.
Using our final result in Eqs. (\ref{x}-\ref{q}), and the
definition of Pauli-Lubanski pseudovector in Eq. (\ref{a-2}), the
space part of $W^\mu$ is obtained as
 \begin{equation}  |\overrightarrow{W}|=\sqrt{C}mcl.
 \end{equation}
By choosing $C=n^{2}$ , if the parameter $l$ is assumed to be the Compton wavelength ($\hbar/mc$), the
magnitude of the spin goes to $n \hbar$. In other words, the
oscillation in Eq. (\ref{x}) occurs at the scale of the Compton
wavelength.
\section{Conclusions}
The main goal of this paper is investigating a systematic method
to find a generalized  Lagrangian for describing the relativistic
 spinning point particle. We know in advance that "mass" and "spin"
constitute two Casimirs of the Poincare group. Therefore, two
constraints $p^2-m^2$ and $W^2+Cm^{4}l^{2}$ should appear in the
Hamiltonian structure of every model introduced for the spinning point
particle. Our idea is based on the fact that the Poincare symmetry is the only
needed tool for constructing the most general form of the required
Lagrangian. In other words, it is possible to read the story from
the end and order the most general form of the Lagrangian of the
required model in such a way that the corresponding
Hamiltonian contains the above constraints as first class
constraints. Then by imposing the Legendre transformation in the
inverse direction one can write the Lagrangian in terms of the
phase space variables.

 However, it is not technically obvious that
one can eliminate the assumed momenta in terms of the
corresponding velocities.
 Fortunately in  our special case of spinning point particle, it is indeed
possible to use the equations of motion of the coordinates to
calculate the quadratic terms appearing in the Hamiltonian in
terms of scalar functions of velocities and coordinates and
use them in the derived form of the Lagrangian (see  Eqs \ref{s-100}and \ref{s-101}).
 Such a transformation    \textit{  from a constrained Hamiltonian
to a singular Lagrangian} is not generally guaranteed to be
possible, while the inverse transformation is normally
performed. Fortunately the problem of spinning
point particle provides a good example of the idea of
"constructing the Lagrangian just by looking on the symmetries".

For a general problem, we know that the generators of a symmetry
group satisfy the corresponding algebra in the framework of the
Poisson brackets on some phase space. Hence, one may consider some
larger or smaller phase space which shows up a representation of
the symmetry algebra among the Poisson brackets.
However, it does not seem a simple problem that under what
conditions one may find a Lagrangian over a configuration space
which lead to the assumed Hamiltonian. In other words, one may
wonder what are the peculiarities of the spinless as well as
the spinning relativistic point particle which enables us to solve
the problem exactly.

We found systematically in this way not only the assumed
Lagrangian of Refs. \cite{kuz} and \cite{7} but also a larger class of the
Lagrangians for the spinning point particle with one more
arbitrary parameter $\alpha$ (see Eq.\ref{Lag}). This parameter is the
initial value of the scalar $k.q$.

In the last two sections we discussed in details the dynamics of our
generalized model in the Lagrangian as well as Hamiltonian formulations. We showed that the complicated
equations of motion support a special oscillatory solution for the spinning particle in the enlarged configuration space.
This oscillatory solution which occurs at the scale of the Compton wavelength of the particle is essential in understanding a classical description for the spin, as showed in Ref. \cite{la}.  Our oscillatory solution again contains the additional parameter $\alpha$, while it reduce to that of Ref. \cite{la}  when $\alpha=0$.

In the framework of the Hamiltonian formalism we showed that the constraint structure consists simply of five first class constraints. Then by fixing the gauges,
as well as the initial conditions, we showed that the oscillatory solution is again supported by the dynamics of the system.

{\bf{Acknowledgements:}} We thank professor A. Deriglazov for valuable comment on first version of this paper. \vskip .3cm


 \end{document}